\documentclass[fleqn,usenatbib]{mnras}

\usepackage{newtxtext,newtxmath}

\usepackage[T1]{fontenc}
\usepackage{ae,aecompl}

\usepackage{graphicx}	
\usepackage{amsmath}	
\usepackage{amssymb}	
\usepackage{comment}
\usepackage{soul}
\usepackage{xcolor}

\newcommand{\cmfast}{\textsc{\small 21CMFAST}}
\newcommand{\cmmc}{\textsc{\small 21CMMC}}

\newcommand{\avenf}{\bar{x}_{\rm HI}}

\newcommand{\Msun}{M_\odot}

\newcommand\lsim{\mathrel{\rlap{\lower4pt\hbox{\hskip1pt$\sim$}}
        \raise1pt\hbox{$<$}}}
\newcommand\gsim{\mathrel{\rlap{\lower4pt\hbox{\hskip1pt$\sim$}}
        \raise1pt\hbox{$>$}}}
\def\myputfigure#1#2#3#4#5%
{\vskip#5pt\makebox[0pt]{\hskip#2in
\includegraphics[width=#3\textwidth]{#1}}\vskip#4pt\hfill}

\title[JWST and 21-cm constraints on the first galaxies]{Properties of reionization-era galaxies from {\it JWST} luminosity functions and 21-cm interferometry}

\author[J. Park et al.]{
Jaehong~Park,$^{1}$\thanks{E-mail: jaehong.park@sns.it (JP)}
Nicolas Gillet$^{1}$
Andrei Mesinger,$^{1}$
and Bradley Greig,$^{2,3}$
\\
$^{1}$Scuola Normale Superiore, Piazza dei Cavalieri 7, I-56126 Pisa, Italy\\
$^{2}$ARC Centre of Excellence for All-Sky Astrophysics in 3 Dimensions (ASTRO 3D), University of Melbourne, VIC 3010, Australia\\
$^{3}$School of Physics, The University of Melbourne, Parkville, VIC 3010, Australia
}

\date{Accepted XXX. Received YYY; in original form ZZZ}

\pubyear{2018}

\hypersetup{draft}
\begin{document}
\label{firstpage}
\pagerange{\pageref{firstpage}--\pageref{lastpage}}
\maketitle

\begin{abstract}
  Next generation observatories will enable us to study the first billion years of our Universe in unprecedented detail.  Foremost among these are 21-cm interferometry with the Hydrogen Epoch of Reionization Arrays (HERA) and the Square Kilometre Array (SKA), and high-$z$ galaxy observations with the James Webb Space Telescope ({\it JWST}).  Taking a basic galaxy model, in which we allow the star formation rates and ionizing escape fractions to have a power-law dependence on halo mass with an exponential turnover below some threshold, we quantify how observations from these instruments can be used to constrain the astrophysics of high-$z$ galaxies.
  For this purpose, we generate mock {\it JWST} LFs, based on two different hydrodynamical cosmological simulations; these have intrinsic luminosity functions (LFs) which turn over at different scales and yet are fully consistent with present-day observations.  We also generate mock 21-cm power spectrum observations, using 1000h observations with SKA1 and a moderate foreground model.
Using only {\it JWST} data, we predict up to a factor of 2-3 improvement (compared with {\it HST}) in the fractional uncertainty of the star formation rate to halo mass relation and the scales at which the LFs peak (i.e. turnover).  Most parameters regulating the UV galaxy properties can be constrained at the level of $\sim 10$\% or better, {\it if} either 
(i) we are able to better characterize systematic lensing uncertainties than currently possible; or (ii) the intrinsic LFs peak at magnitudes brighter than $M_{\rm UV} \lsim -13$. Otherwise, improvement over {\it HST}-based inference is modest. When combining with upcoming 21-cm observations, we are able to significantly mitigate degeneracies, and constrain all of our astrophysical parameters, even for our most pessimistic assumptions about upcoming {\it JWST} LFs.  The 21-cm observations also result in an order of magnitude improvement in constraints on the EoR history.
\end{abstract}




%
%
\section{Introduction}\label{sec:intro}

Recent years have witnessed remarkable progress in understanding the timing of the epoch of reionization (EoR).  Aided primarily by high-redshift QSO spectra \citep[e.g.][]{Mortlock2011, McGreer2015,Banados2018} and the optical depth to the cosmic microwave background (CMB) \citep[e.g.][]{Planck2016,Planck2018}, we can estimate that the mid-point of the EoR (when the volume-averaged neutral fraction was $\avenf = 0.5$) was around $z\sim7.5 \pm 1$, (e.g. \citealt{Mitra2015, Planck2016, Greig&Mesinger2017, Price2018, Gorce2018}) with a maximum of a few percent of the IGM remaining neutral by $z=6$ (\citealt{McGreer2015}; although the final overlap stages can extend to $z\sim5$--6; \citealp{Lidz2007, Mesinger2010, Keating2019}).

The next few years will see us moving away from putting points on the $\avenf$ vs. $z$ plane, towards a deeper understanding of the galaxies that are responsible for the EoR.  This will primarily be enabled by two ground-breaking observations: (i) near infrared high-$z$ galaxy studies with the James Webb Space Telescope ({\it JWST}; \citealp{Gardner2006}) and (ii) measurements of the 3D structure of the EoR with next-generation 21-cm interferometers like Hydrogen Epoch of Reionization Array (HERA\footnote{http://reionization.org}; \citealp{DeBoer2017}) and Square Kilometre Array(SKA\footnote{https://astronomers.skatelescope.org}; \citealp{Mellema2013,Koopmans2015}).

Although JWST will enable resolved spectroscopy of high-$z$ galaxies, such detailed studies will be limited to relatively bright and rare objects \citep[e.g.][]{Stark2016,Shapley2017,Williams2018,Chevallard2019}.
The bulk of the high-$z$ galaxy population will be studied primarily by counting the number per volume which fall in a given non-ionizing UV magnitude bin, the so-called rest-frame UV luminosity functions (UV LFs).  {\it JWST} should extend our knowledge of high-z LFs by pushing one to two magnitudes deeper than current observations with Hubble \citep[e.g.][]{Salvaterra2011,Dayal2013,Shimizu2014,O'Shea2015,Finkelstein2016,Wilkins2017,Cowley2018,Tacchella2018,Williams2018,Yung2019}. This will allow us to push blank field LFs to magnitudes fainter than $M_{\rm UV} \gsim -17$; such faint magnitudes are currently accessible only through cluster lensing, and are thus susceptible to large systematic uncertainties including lens modeling and completeness corrections (e.g. \citealt{Bouwens2016_LF6,Livermore2017_LF,Atek2018,Ishigaki2018}).

On the other hand, the 21-cm line from neutral hydrogen will enable us to map the intergalactic medium (IGM) on large-scales, during the first billion years.  From these large-scale 21-cm structures, we can {\it indirectly} infer average properties of high-redshift galaxies, albeit with some degeneracies \citep[e.g.][]{McQuinn2007,Pober2015,21CMMC,Greig2017, Ross2019}. These properties include the stellar mass fraction, the gas fraction, the star formation rate, the escape fraction, X-ray luminosities, etc.

In \citet{Park2019} we showed that high-$z$ LFs and 21-cm interferometry are complementary observations, helping us nail down the properties of high-$z$ galaxies, and ameliorating the degeneracies present when each is considered separately.  We used current LF observations obtained with the {\it Hubble} telescope, combining them with a mock 21-cm observation from a 1000h integration with the HERA instrument.  {\it In this work, we quantify the additional constraints on high-z galaxy properties available with deeper LF observations, such as might be expected from {\it JWST}.}

This paper is organized as follows. In \S~\ref{sec:data} we describe our mock LF and 21-cm observations. Then, we show the corresponding constraints on astrophysical parameters in \S~\ref{sec:results}. In \S~\ref{sec:conc}, we summarize our results. We assume a standard ${\rm \Lambda}$CDM cosmology based on {\it Planck} 2016  \citep{PlanckXIII}: ($h$, $\Omega_{\rm m}$, $\Omega_{\rm b}$, $\Omega_{\Lambda}$, $\sigma_{8}$, $n_{\rm s}$)=(0.678, 0.308, 0.0484, 0.692, 0.815, 0.968).  Unless stated otherwise, we quote all quantities in comoving units, and  when we refer to the UV magnitude, this corresponds to the rest-frame 1500{\AA}  AB magnitude.

%
%
\section{Data sets}\label{sec:data}

As in \cite{Park2019}, we compute the likelihood of our model parameters using two main data sets: the rest-frame UV LFs at high-$z$ and mock 21-cm power spectra (PS) measurements.  The main difference in this work is that instead of using current LFs observations from {\it Hubble}, we use deeper mock LFs, roughly corresponding to what we should get with {\it JWST}.

In addition to these two mock data sets, we also include in our likelihood calculation the two most robust constraints on EoR timing currently available:
 (i) the electron scattering optical depth to the CMB, $\tau_{\rm e} = 0.058 \pm 0.012 (1\sigma)$ from \citep{Planck2016}; and (ii) the upper limit of the neutral fraction, $\avenf < 0.06 + 0.05 (1\sigma)$ at $z=5.9$ from the fraction of dark pixels in QSO spectra \citep{McGreer2015}. These EoR timing measurements allow a rough estimate of the ionizing escape fraction, when combined also with the observed LFs \citep[e.g.][]{Kuhlen2012, Mitra2013, Mitra2015, Robertson2013, Robertson2015,Price2016}, but become superfluous when 21-cm observations become available \citep{Park2019}.


%
%
\subsection{Mock JWST LFs}\label{sec:LFs}

\begin{figure}
    \includegraphics[width=\columnwidth]{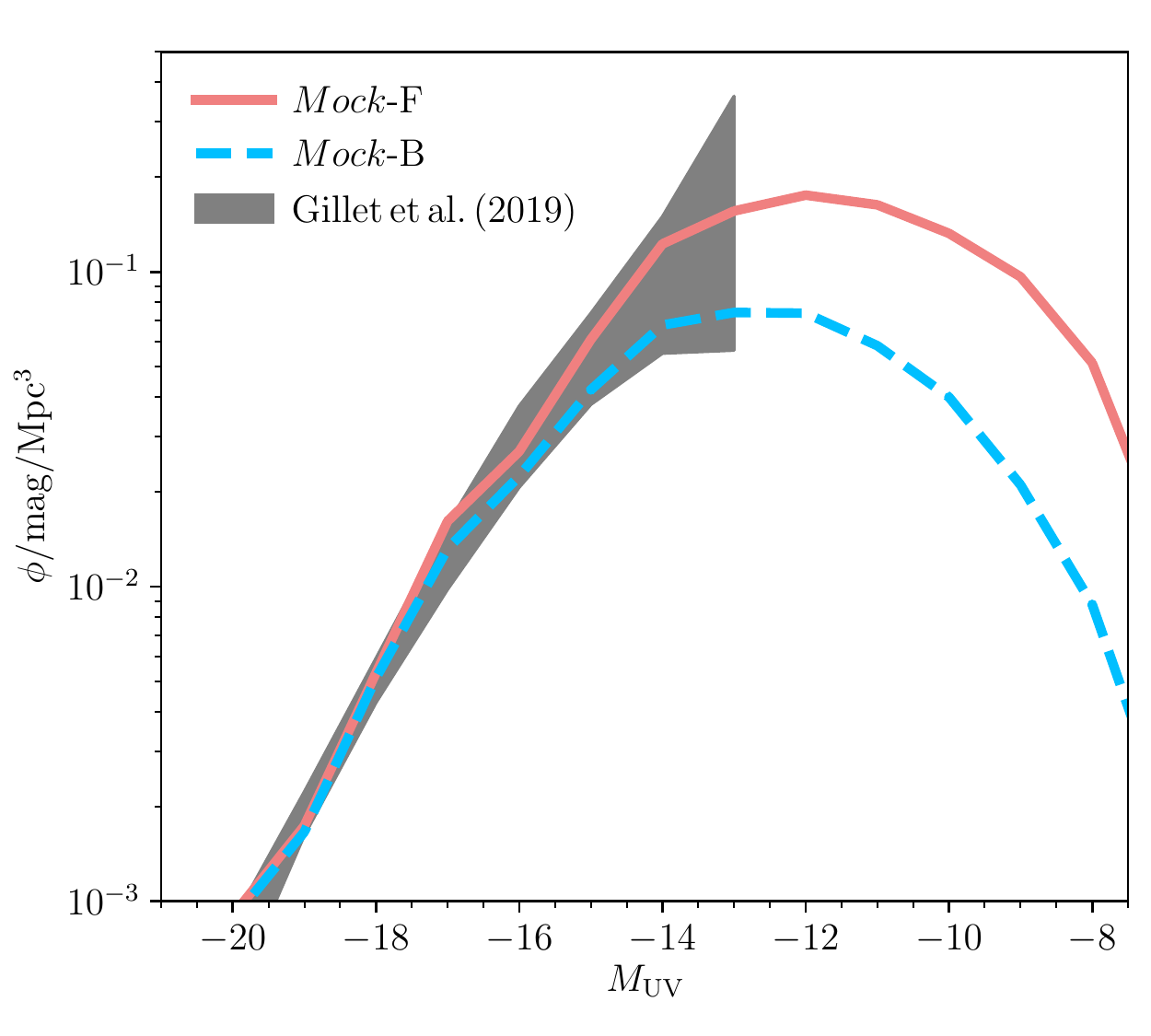}
    \caption{ Simulated luminosity functions at $z=6$.  These simulations were used to build the two sets of mock {\it JWST} LFs shown in Fig. 2.  These were chosen because they agree with current constraints (the 68\% C.L. from \citealt{Gillet2019} are denoted with the shaded area) but are very different at the ultra faint end.}
\label{fig:LF_EMMA}
\end{figure}

Our mock LFs are taken from the GAlaxy Formation For the EoR (GAFFER) simulation suite (Gillet et al. in prep).  
GAFFER is comprised of $\sim$800 fully-coupled hydro-radiative transfer cosmological simulations aiming to characterize the growth of dwarf galaxies during the EoR.  Using the numerical code EMMA \citep{Aubert2015}, we vary five astrophysical/numerical parameters governing galaxy formation: the star formation efficiency, the ISM over-density threshold for star formation, the supernova feedback efficiency, the sub-grid ionizing escape fraction, and the mass of the numerical star particle.
For more details on how these parameters affect the star formation and feedback models in EMMA, we refer the reader to \citet{Deparis2016, Deparis2019} and Deparis et al in prep.
Most of the simulation boxes are 10Mpc on a side, with halos above $\sim2 \times 10^8 M_{\odot}$ being resolved with $>100$ dark matter particles.

\begin{figure*}
\vspace{-1\baselineskip}
{
 \includegraphics[width=\textwidth]{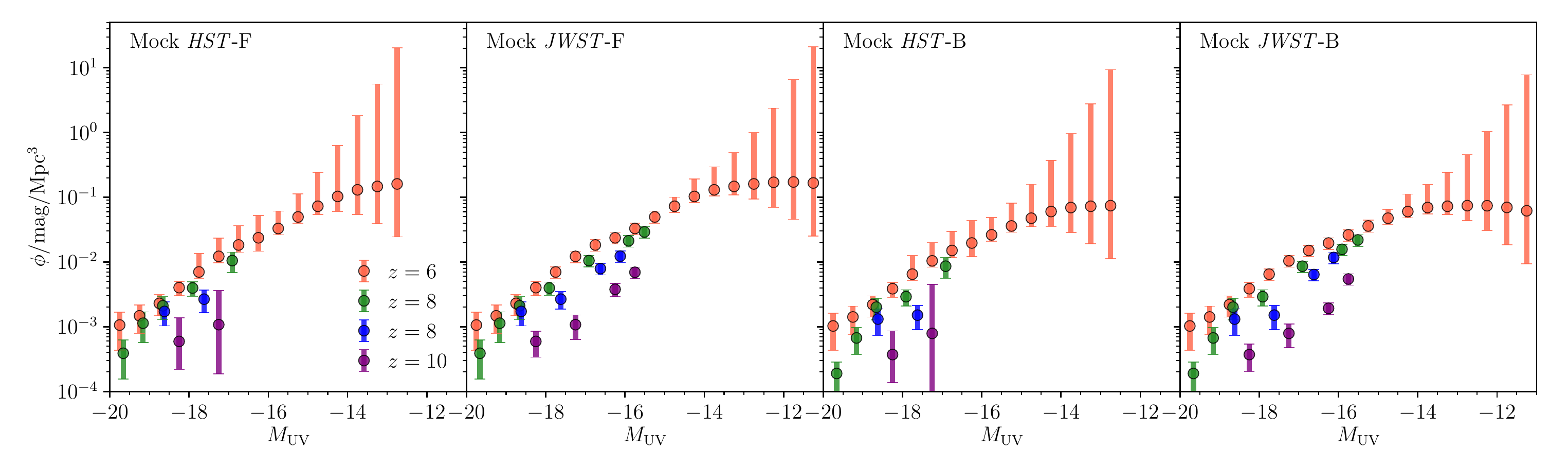}
}
\caption{
  {\it HST} and {\it JWST} LF mock observations used for parameter recovery.
    LFs corresponding to the simulation with a turnover at brighter (fainter) magnitudes are denoted with ``-B'' (``-F'').
\label{fig:LFs}
}
\vspace{-0.5\baselineskip}
\end{figure*}

For each simulation, we compute the corresponding LFs by assuming a constant conversion factor between a galaxy's star formation rate (averaged over the previous 10 Myr)\footnote{We note that had we chosen a longer time-frame over which to average the SFR, the drop at faint magnitudes ($M_{\rm UV}$>-10) seen in Fig. 1 would be less steep; however, this does not have a notable effect on the observable luminosity range.} and its 1500{\AA} UV luminosity:  $\dot{M_{\ast}} = \mathcal{K}_{\rm UV} \times L_{\rm UV}$, with $\mathcal{K}_{\rm UV} = 1.15 \times 10^{-28} {\rm \Msun\,yr^{-1}/erg\,s^{-1}\,Hz^{-1}}$ consistent with a Salpeter IMF with a $\sim$10\% solar metalicity (c.f. \citealt{Kennicutt1998, Madau&Dickinson2014}).  We do not model dust; thus our mock LFs would correspond to dust-corrected ones.  However, since we concern ourselves with faint galaxies which dominate the photon budget during reionization, dust is unlikely to require a large correction over our magnitude range \citep [e.g.][]{Finkelstein2012, Dunlop2013, Bouwens2015, Cullen2017, Wilkins2017, Yung2019, Ma2019, Vogelsberger2019}.

From the GAFFER simulation suite, we select two simulations to use as mock {\it JWST} LFs.  These two simulations have LFs which are both within 1 $\sigma$ of current observational constraints (e.g. \citealt{Gillet2019}; see Fig. \ref{fig:LF_EMMA}) but were chosen to have different behavior at the faint end, due to different strengths of SNe and photo-heating feedback from reionization (Gillet et al. in prep).  Detecting a turnover at brighter magnitudes with {\it JWST} would be easier than detecting it at fainter magnitudes; thus we take these two models to bracket the expected range.  Below we denote mock observations based on the simulation with a turnover at brighter magnitudes with the suffix ``-B'', and those based on the simulation with a turnover at fainter magnitudes with the suffix ``-F''.

Mock observations are constructed from the galaxy number densities, $\phi(M_{\rm uv})$, in these two simulations.  For consistency, we use the same galaxy number counts to make both {\it JWST} and {\it HST} mock LFs, varying only the uncertainties (by construction, the {\it HST} mocks are entirely consistent with current {\it HST} observations).  The uncertainties for {\it HST} are the same ones we used in \cite{Park2019}, allowing for a direct comparison.  These uncertainties were based on the observational data from
\cite{Bouwens2016_LF6} for redshift 6, \cite{Bouwens2015_LF4-10} for redshifts 7 and 8, and \cite{Oesch2018} for redshift 10. 
The same magnitude bins are used. In each bin, the error is evaluated as the maximum of the observational uncertainty ($\sigma_{\rm{OBS}}$) and the Poisson error in the bin due to the finite simulation volume of the simulations ($\sigma_{\rm P}$). Over most of the range of interest, specifically $M_{\rm UV} > -18$, the observational uncertainties dominate over the cosmic variance.  Moreover, we enforce that the uncertainty has to be greater than 20\% of the galaxy density ($\sigma \geq 0.2 \phi$, taking this to be a systematic ``floor'' (R. Bouwens, private communication).

The mock {\it JWST} LFs are then built by extending the {\it HST} LFs 1.5 magnitudes deeper.  Although in principle, error bars should be computed considering specific volumes of specific observational programs, simply extending the {\it HST} LFs by 1.5 mags is a reasonable approximation for what is achievable with {\it JWST} (\citealp{Finkelstein2016}; R. Bouwens and P. Oesch, private communication).
Specifically, we take:
\begin{align}
  \sigma_{{\it JWST}}(M_{\rm UV}) =
  \left\{
  \begin{array}{lll}
    \sigma_{\it HST}(M_{\rm UV} - 1.5 ) ~ & {\rm if} ~  M_{\rm UV} > -14.5\\
    \sigma_{\it HST}(M_{\rm UV}) ~ & {\rm if} ~  M_{\rm UV} < -18\\
    0.2\phi(M_{\rm UV})  ~ & {\rm if} ~  -18 < M_{\rm UV} < -14.5
  \end{array}
  \right.
\end{align}
with the Poisson-dominated bright-end taken to have the same errors as are currently available for {\it HST}, and the intermediate regime having 20\% systematic errors, similar to what is available currently from {\it HST} programs.

The resulting mock LFs are shown in Fig. \ref{fig:LFs}, for both {\it HST} and {\it JWST} error bars, as well as for both of our simulations.  As mentioned previously, LFs which have an intrinsic turnover at fainter (brighter) magnitudes are denoted with the qualifiers ``-F" (``-B").

%
%
\subsection{Mock 21-cm signal}\label{sec:21cm}

We create a mock cosmic 21-cm signal using the public code {\cmfast}\footnote{\url{https://github.com/andreimesinger/21cmFAST}}. {\cmfast} \citep{Mesinger2007,21cmfast} generates the evolved density and corresponding peculiar velocity fields by applying second order LPT \citep[e.g.][]{Scoccimarro1998} on a high-resolution realization of a Gaussian random field.  Then, {\cmfast} estimates the ionization field from the density field using an excursion-set approach (e.g. \citealt{Furlanetto2004}), while the spin temperature evolution is computed by integrating the cosmic X-ray and soft UV backgrounds back along the lightcone for each simulation cell. We use the latest version introduced in \citet{Park2019}, allowing us to tie the galactic radiation sources to the corresponding UV LFs (c.f. \url{http://homepage.sns.it/mesinger/Videos/parameter_variation.mp4}). Here we briefly summarize the free parameters in the model; for more details on the simulation and the astrophysical parameters, readers are referred to \cite{Park2019}.   

We assume the average properties of high-$z$ galaxies depend on their host dark matter halo mass \citep[e.g.][]{Behroozi2015,Sun&Furlanetto2016,Dayal&Ferrara2018,Salcido2019}.
Specifically, we parametrize the typical stellar mass of galaxies with a power law dependence on the total halo mass, $M_h$:
\begin{equation}\label{eq:F_STAR}
M_{\ast}(M_{\rm h}) = f_{\ast,10} \left( \frac{M_{\rm h}} {10^{10}{\rm M}_{\sun}} \right)^{\alpha_{\ast}} \left(\frac{\Omega_b}{\Omega_m}\right) M_h
\end{equation}
where $f_{\rm \ast,10}$ is the normalization (i.e. the fraction of galactic baryons in stars for halos with a mass of $10^{10} {\rm M}_\odot$) and $\alpha_{\ast}$ is the power-law index. Then, the star formation rate (SFR) is defined as 
\begin{equation}\label{eq:SFR}
  \dot{M_{\ast}}(M_{\rm h},z) =  \frac{M_{\ast}}{t_\ast H(z)^{-1}},
\end{equation}
where $H(z)^{-1}$ is the Hubble time and $t_{\ast}$ is a free parameter regulating the star formation time-scale.

Similarly we define the ionizing UV escape fraction as
\begin{equation}\label{eq:F_ESC}
f_{\rm esc}(M_{\rm h}) = f_{{\rm esc, 10}}\left( \frac{M_{\rm h}}{10^{10}{\rm M}_{\sun}}\right)^{\alpha_{\rm esc}},
\end{equation}
where $f_{{\rm esc, 10}}$ is the normalization of the escape fraction and $\alpha_{\rm esc}$ is a power-law index.

Since small halos are unable to host star-forming galaxies due to their limited gas reservoir from inefficient cooling and/or feedback
 \citep[e.g.][]{Shapiro1994,Giroux1994,Hui1997,Barkana&Loeb2001,Springel&Hernquist2003,Okamoto2008,Mesinger2008,Sobacchi2013a}, we introduce a duty cycle quantifying the fraction of halos which host galaxies via
\begin{equation}\label{eq:DC}
f_{\rm duty}(M_{\rm h}) = \exp\left( - \frac{M_{\rm turn}}{M_{\rm h}}\right),
\end{equation}
Here $M_{\rm turn}$ is a characteristic mass below which the fraction of halos hosting stars/galaxies exponentially decreases.  For reference, $M_{\rm turn} \sim 10^8 \Msun$ at $z\sim10$ for a virial temperature of $10^4$ K (corresponding to the atomic cooling threshold).

The corresponding rest-frame UV LFs are calculated as:
\begin{equation}\label{eq:LF}
\phi(M_{\rm UV}) = \left[ f_{\rm duty} \frac{{\rm d}n}{{\rm d}M_{\rm h}} \right] \left|\frac{{\rm d}M_{\rm h}}{{\rm d}M_{\rm UV}}\right| ~ ,
\end{equation}
where ${\rm d}n/{\rm d}M_{\rm h}$ is the halo mass function. To calculate the ${\rm d}M_{\rm h}/{\rm d}M_{\rm UV}$ term, we assume a linear dependence of the 1500 \AA\  UV luminosity to the star formation rate:  $\dot{M_{\ast}}(M_{\rm h}, z) = \mathcal{K}_{\rm UV} \times L_{\rm UV}$, just as we did when constructing the mock LFs from the GAFFER simulations.

We thus have six free parameters which regulate the emission of UV photons: $f_{\ast,10}$, $\alpha_{\ast}$, $f_{{\rm esc, 10}}$, $\alpha_{\rm esc}$, $M_{\rm turn}$ and $t_{\ast}$.  We introduce two additional parameters to characterize the X-ray emission of high-$z$ galaxies, $L_{\rm X<2\,keV}/{\rm SFR}$ and $E_0$, which we describe below.

It is expected that X-rays, through their long mean free paths, are a dominant source of heat in the neutral IGM, outside of the HII regions which surround the nascent galaxies
\citep[e.g.][]{Pritchard&Furlanetto2007,McQuinn&O'Leary2012,Mesinger2013, Madau&Fragos2017, Eide2018}.
          {\cmfast} computes the angle-averaged specific X-ray intensity  (in units of ${\rm erg\,s^{-1}\,keV^{-1}\,cm^{-2}\,sr^{-1}}$) in each simulation cell at a given spatial position and redshift. We parametrize the typical emerging X-ray SED of high-$z$ galaxies via their integrated soft-band ($< 2{\rm keV}$) luminosity per SFR (in units of ${\rm erg\,s^{-1}\,M_{\sun}^{-1}\,yr}$),
\begin{equation}\label{eq:soft_X-ray}
L_{\rm X<2\,keV}/{\rm SFR} = \int_{E_0}^{2\,{\rm keV}}{\rm d}E_{\rm e}\, L_{\rm X}/{\rm SFR},
\end{equation}
where $L_{\rm X}/{\rm SFR}$ is is the specific X-ray luminosity per unit star formation escaping the host galaxies in units of ${\rm erg\,s^{-1}\,keV^{-1}\,M_{\sun}^{-1}\,yr}$, taken here to be a power law with energy index $\alpha_X=1$ (e.g. \citealt{Fragos2013, Mineo2012, Das2017}) and $E_0$ is an additional free parameter corresponding to the X-ray energy threshold below which photons are absorbed inside the host galaxies, never managing to escape and heat the IGM.

We compute a mock observation from a simulation box of $500\,{\rm Mpc}$ on a side with a $256^3$ grid, downsampled from $1024^3$ initial conditions.  Our default astrophysical parameters used for the mock simulation are listed in Table 1; these parameters are consistent with the mock UV LFs as shown below and discussed in Gillet at al. in prep.

From the light-cone of this simulation, we compute the 3D power spectra in 12 segments, sliced along the redshift/frequency axis in equal comoving volumes. As in \citet{Park2019}, we compute the thermal and cosmic variance noise on the power spectrum at each redshift using the public code \textsc{21cmsense}\footnote{\url{https://github.com/jpober/21cmSense}}  \citep{Pober2013,Pober2014}. For this, we assume the `moderate' foreground removal strategy from \citet{Pober2014} which restricts the computation of the 21cm PS to modes outside of the foreground `wedge'. Further, this assumes coherent summation over redundant baselines in order to reduce thermal noise \citep{Parsons2012}.
For this work, we assume a single 1000hr tracked scan with the SKA. We model the SKA using the recent SKA System Baseline Design document\footnote{http://astronomers.skatelescope.org/wp-content/uploads/2016/09/SKA-TEL-SKO-0000422\textunderscore 02\textunderscore SKA1\textunderscore LowConfigurationCoordinates-1.pdf}

%
%
\section{Results}\label{sec:results}

\begin{figure*}
\vspace{-1\baselineskip}
{
 \includegraphics[trim={0 5cm 0 0},clip,width=\textwidth]{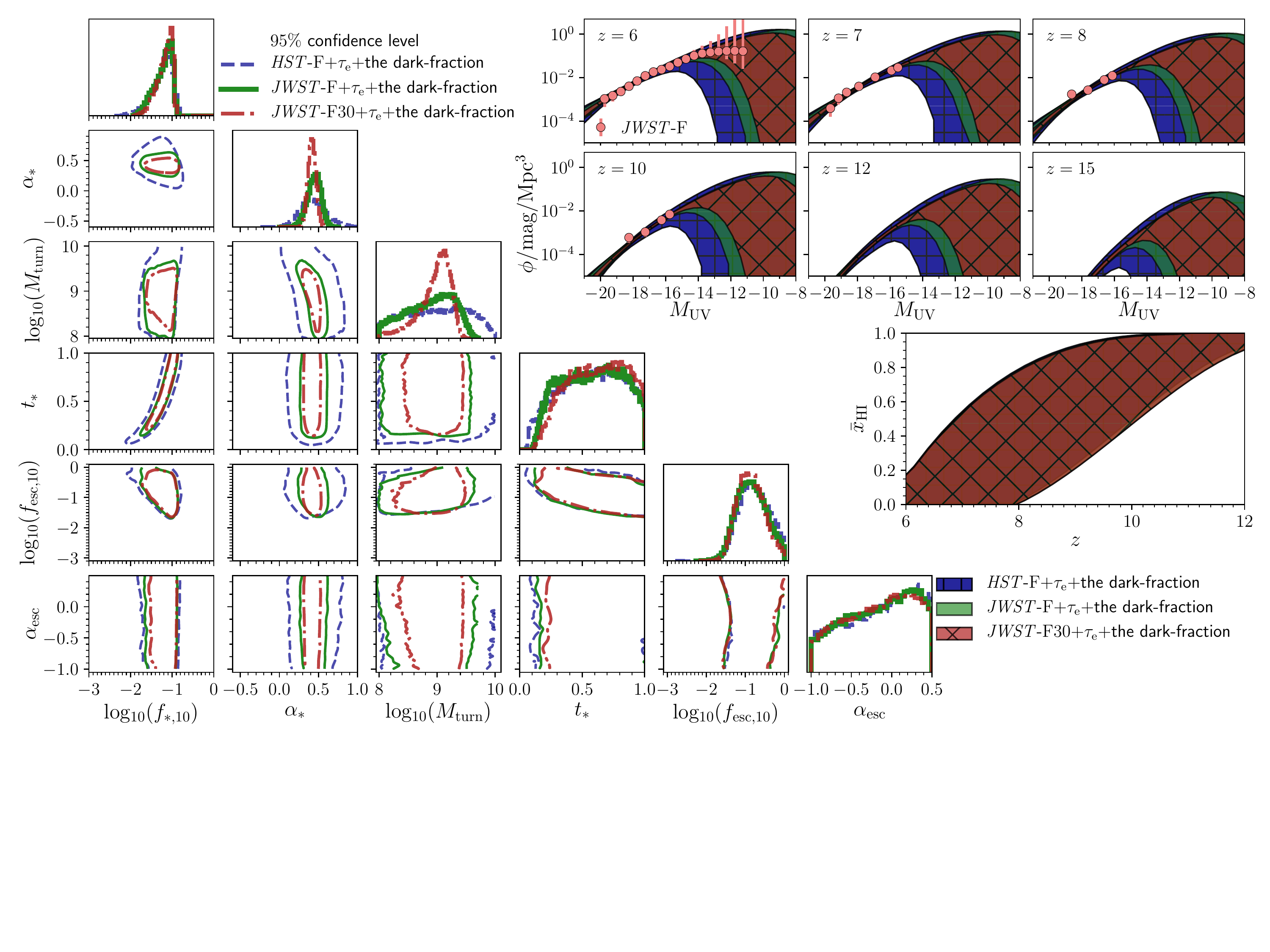}
}
\caption{
 Corner plot showing parameter constrains for the mock UV LFs (see legend): 1D marginalized PDFs and 2D marginalized joint posterior distributions are shown along the diagonal and in the bottom left corner, respectively. Blue dashed lines, green solid lines and brown dot-dashed lines represent $95$ per cent confidence levels for constraints using data sets of the mock ${\it HST}$-F, the mock ${\it JWST}$-F and the mock {\it JWST}-F30, respectively. Top-right panels: Recovered $95$ per cent confidence levels of the LFs corresponding to the posterior of our model. Shaded regions with the cross hatch (blue, `$+$'), shaded regions (green) and shaded regions with  `$\times$' hatch (brown) represent constraints using the mock ${\it HST}$-F, the mock ${\it JWST}$-F and the mock {\it JWST}-F30, respectively. Middle-right: Corresponding constraints on the global evolution of the IGM neutral fraction, $x_{\ion{H}{I}}(z)$ with the same legends (note that these almost entirely overlap, highlighting that improved LF constraints will not aid in nailing down the EoR history, provided the escape fraction is allowed to be a free functional). Note that together with the listed data sets we also use (i) the electron scattering optical depth to the CMB from \protect\cite{Planck2016}; and (ii) the upper limit of the neutral fraction at $z=5.9$ from the dark fraction of pixels in QSO spectra \protect\citep{McGreer2015}.
\label{fig:corner_LFs1}
}
\vspace{-0.5\baselineskip}
\end{figure*}

We combine the above mentioned data sets within a fully Bayesian framework, obtaining parameter constraints with the public Monte Carlo Markov Chain (MCMC) sampler of 3D EoR/CD simulations, {\cmmc}\footnote{\url{https://github.com/BradGreig/21CMMC}} \citep{21CMMC,Greig2017,Greig2018}. 
At each parameter sample, {\cmmc} computes the corresponding 21-cm PS, UV LFs, and reionizaton history, comparing them with our data sets using a $\chi^2$ likelihood and a flat prior over the parameter ranges shown in the figures below. The likelihoods for each data set are multiplied together when computing the posterior. For the 21-cm PS, we include a $20\%$ Gaussian error on the PS in each bin to account for simulation inaccuracy (e.g. \citealt{Zahn2011}), adding it in quadrature with the sample variance.
For all runs, we include the additional EoR timing constraints mentioned above: (i) the electron scattering optical depth to the CMB $\tau_{\rm e} = 0.058 \pm 0.012 (1\sigma)$ from \protect\cite{Planck2016}; and (ii) the upper limit of the neutral fraction $\avenf < 0.06 + 0.05\,(1\sigma)$ at $z=5.9$ from \cite{McGreer2015}.

%
%
\subsection{Constraints using LFs without 21-cm}\label{sec:result_LFs}

%
%
\subsubsection{Assuming an intrisic turnover at fainter magnitudes}\label{sec:result_LFs_faint}
In Fig.\ref{fig:corner_LFs1}, we show constraints on our six astrophysical parameters describing the UV emission of galaxies: $f_{\ast,10}$, $\alpha_{\ast}$, $f_{{\rm esc, 10}}$, $\alpha_{\rm esc}$, $M_{\rm turn}$ and $t_{\ast}$, constructed using the ``faint end'' turnover LFs, for both {\it HST} and {\it JWST}.  As discussed previously, if the turnover is at faint magnitudes, it is more difficult to be detectable even with {\it JWST}; therefore the ``-F'' LFs can be considered the ``pessimistic'' scenario.

The marginalized posteriors are shown in the corner plot on the left side, while the corresponding recovered UV LFs are shown in the upper right, with the EoR history in the middle right.  As noted in the legend, blue / green lines and shaded areas denote posteriors constructed using {\it HST}-F / {\it JWST}-F data sets.  All data sets additionally include $\tau_{\rm e}$ and the dark fraction measurements.  The marginalized 1D constraints are also written in Table \ref{Table:recovered_parameters}.

Using {\it HST}-F LFs, we recover the trends already noted in in \cite{Park2019}.  Although here we use a mock {\it HST}-F observation to directly compare against the {\it JWST}-F forecast, the mock {\it HST}-F is by construction consistent with current observations (c.f. Fig. \ref{fig:LF_EMMA}), and so the agreement with \citet{Park2019} is understandable.  Most notably, we find that {\it HST}-F LFs are unable to constrain the flattening/turnover scale, $M_{\rm turn}$, encoding the halo mass below which star formation becomes inefficient.  They only provide an upper limit, ruling out ${\rm log}_{10}(M_{\rm turn}) \lesssim 9.88$ at $95$ per cent confidence level.  The constraints on the scaling of the stellar mass with halo mass is reasonable, with a fractional uncertainty in the relevant parameters of order tens of percent (c.f. Table 1).\footnote{A careful reader can note that the recovered fractional uncertainty on  $\alpha_{\ast}$ is a factor of two larger than quoted in \citet{Park2019}.  This is due to the fact that our mock observation comes from a fairly small simulation box, 10 Mpc on a side.  The resulting Poisson noise for the brightest galaxies is larger than was quoted in the \citet{Bouwens2016_LF6} observations that were used in \citet{Park2019}, resulting in weaker recovery on the halo mass scaling of the stellar mass.}
The ionizing escape fraction is only poorly constrained, with the normalization parameter $f_{\rm esc,10}$ having a $1\sigma$ fractional uncertainty of $\sim 50$ per cent, while its dependence on halo mass is completely unconstrained (as evidenced by the flat marginalized PDF over $\alpha_{\rm esc}$, consistent with our priors).

\begin{table*}
\begin{center}
\caption{
         Summary of the median recovered values with $1\,\sigma$ errors for the eight free parameters, obtained from our MCMC procedure for each combination of data sets listed below. Note that together with the listed data sets we also use (i) the electron scattering optical depth to the CMB from \protect\cite{Planck2016}; and (ii) the upper limit of the neutral fraction at $z=5.9$ from the dark fraction of pixels in QSO spectra \protect\citep{McGreer2015}. We note that the fiducial values are used for generating the mock 21-cm signal; LFs are taken independently from the GAFFER simulations.}
\begin{tabular} {ccccccccc}
\\
\hline\\[-3.0mm]
               &  &  &  &  Parameters  &  &  &       \\[1mm]
               & ${\rm log_{10}}(f_{\ast,10})$ & $\alpha_{\ast}$ & ${\rm log_{10}}(f_{\rm esc,10})$ & $\alpha_{\rm esc}$ & ${\rm log_{10}}(M_{\rm turn})$ & $t_{\ast}$ & ${\rm log_{10}}\left(\frac{L_{{\rm X}<2{\rm keV}}}{\rm SFR}\right)$ & $E_0$   \\[1mm]
               &  &  &  &  & $[{\rm M_{\sun}}]$  &   & $[{\rm erg\,s^{-1}\,M_{\sun}^{-1}\,yr}]$ &  $[{\rm keV}]$   \\[1.5mm] \hline \\[-2.5mm]
  Fiducial values (21-cm only)  & $-1.155$ & $0.38$ & $-1.155$ & $-0.20$ & $9.00$  & $0.6$ & $40.50$ &  $0.50$   \\[1.5mm] \hline\\[-2.5mm]              
  {\it HST}-F      & $-1.19^{+0.18}_{-0.31}$ &  $0.44^{+0.18}_{-0.15}$ &  $-0.79^{+0.43}_{-0.38}$  &  $-0.09^{+0.42}_{-0.57}$   &   $9.02^{+0.56}_{-0.63}$   &  $0.58^{+0.26}_{-0.29}$   &    .  &  . \\[1.5mm]
 {\it JWST}-F     & $-1.17^{+0.16}_{-0.28}$ &  $0.45^{+0.08}_{-0.09}$ &  $-0.85^{+0.41}_{-0.37}$  &  $-0.11^{+0.42}_{-0.54}$   &   $8.92^{+0.38}_{-0.52}$   &  $0.59^{+0.26}_{-0.28}$   &   .  &   .\\[1.5mm]
 {\it JWST}-F30     & $-1.15^{+0.14}_{-0.23}$ &  $0.42^{+0.05}_{-0.05}$ &  $-0.87^{+0.37}_{-0.33}$  &  $-0.12^{+0.41}_{-0.53}$   &   $9.02^{+0.19}_{-0.33}$   &  $0.63^{+0.24}_{-0.26}$   &    .  &   . \\[1.5mm]
   {\it HST}-B      & $-1.29^{+0.19}_{-0.29}$ &  $0.36^{+0.14}_{-0.12}$ &  $-0.64^{+0.38}_{-0.39}$  &  $-0.14^{+0.44}_{-0.55}$   &   $9.11^{+0.53}_{-0.65}$   &  $0.59^{+0.26}_{-0.27}$   &    .  &  . \\[1.5mm]
 {\it JWST}-B     & $-1.24^{+0.16}_{-0.26}$ &  $0.38^{+0.09}_{-0.09}$ &  $-0.62^{+0.33}_{-0.33}$  &  $-0.15^{+0.45}_{-0.53}$   &   $9.40^{+0.18}_{-0.36}$   &  $0.61^{+0.25}_{-0.27}$   &   .  &   .\\[1.5mm]
21-cm $+$ {\it JWST}-F   &  $-1.19^{+0.12}_{-0.15}$ &  $0.44^{+0.07}_{-0.06}$  &  $-1.16^{+0.16}_{-0.13}$  &  $-0.21^{+0.10}_{-0.10}$  &   $8.92^{+0.11}_{-0.10}$   &  $0.57^{+0.17}_{-0.16}$   &   $40.49^{+0.04}_{-0.04}$   &   $0.50^{+0.02}_{-0.02}$\\[1.5mm]
\hline
  
\end{tabular}
\label{Table:recovered_parameters}
\end{center}
\end{table*}

Considering the {\it JWST}-F LFs, we note that the constraints are not very different for the escape function parameters (as is expected since we are not directly adding information on the ionizing photon budget).  However, the recovery of parameters describing star formation is (modestly) improved.  Specifically, we note that the $1\sigma$ fractional uncertainty for $\alpha_{\ast}$ is reduced by a factor of 2. This is because the reduced errors of the mock {\it JWST} LFs tighten the slope of the LFs (see Fig. \ref{fig:LFs}). Together with the reduced errors, the extended faint-end provides additional information on the abundance of faint galaxies, which translates to a somewhat tighter upper limit on ${\rm log}_{10}(M_{\rm turn}) \lesssim 9.53$ at $95$ per cent confidence level. This improvement is more notable when looking at the corresponding recovered LFs in the upper right panels.  We see explicitly that our mock {\it JWST}-F LFs allow us to rule out models which predict a turnover at $M_{\rm UV} < -13$.

%
%
\subsubsection{Would reduced observational uncertainties improve constraints?}\label{sec:result_JWST-F30}

In the previous section, we noted that {\it if} the intrinsic LFs turn over at faint magnitudes ($M_{\rm UV}\gsim -12$), {\it JWST} LFs will only modestly improve on our current knowledge of average galaxy properties, as obtained with {\it HST} LFs. The largest improvement comes in the form of improved constraints on $\alpha_\ast$ and a somewhat tighter upper limit on the turnover scale.

Here we consider a more "optimistic" {\it JWST}-F forecast, labeled {\it JWST}-F30. This forecast is based on the same intrinsic LFs, ``-F'', but we assume that uncertainties can be reduced, e.g. due to an improved understanding of the dominant systematic uncertainties.  To illustrate this, we simply reduce the errors of each LF bin to $30\%$ of their fiducial values, discussed previously, keeping the 20\% minimum error.  In other words, in each magnitude bin we take $\sigma_{\it JWST-{\rm F30}} = {\rm max}[\sigma_{\it JWST-{\rm F}}, 0.2 \phi]$. 

\begin{figure}
\vspace{-1\baselineskip}
{
 \includegraphics[width=8cm]{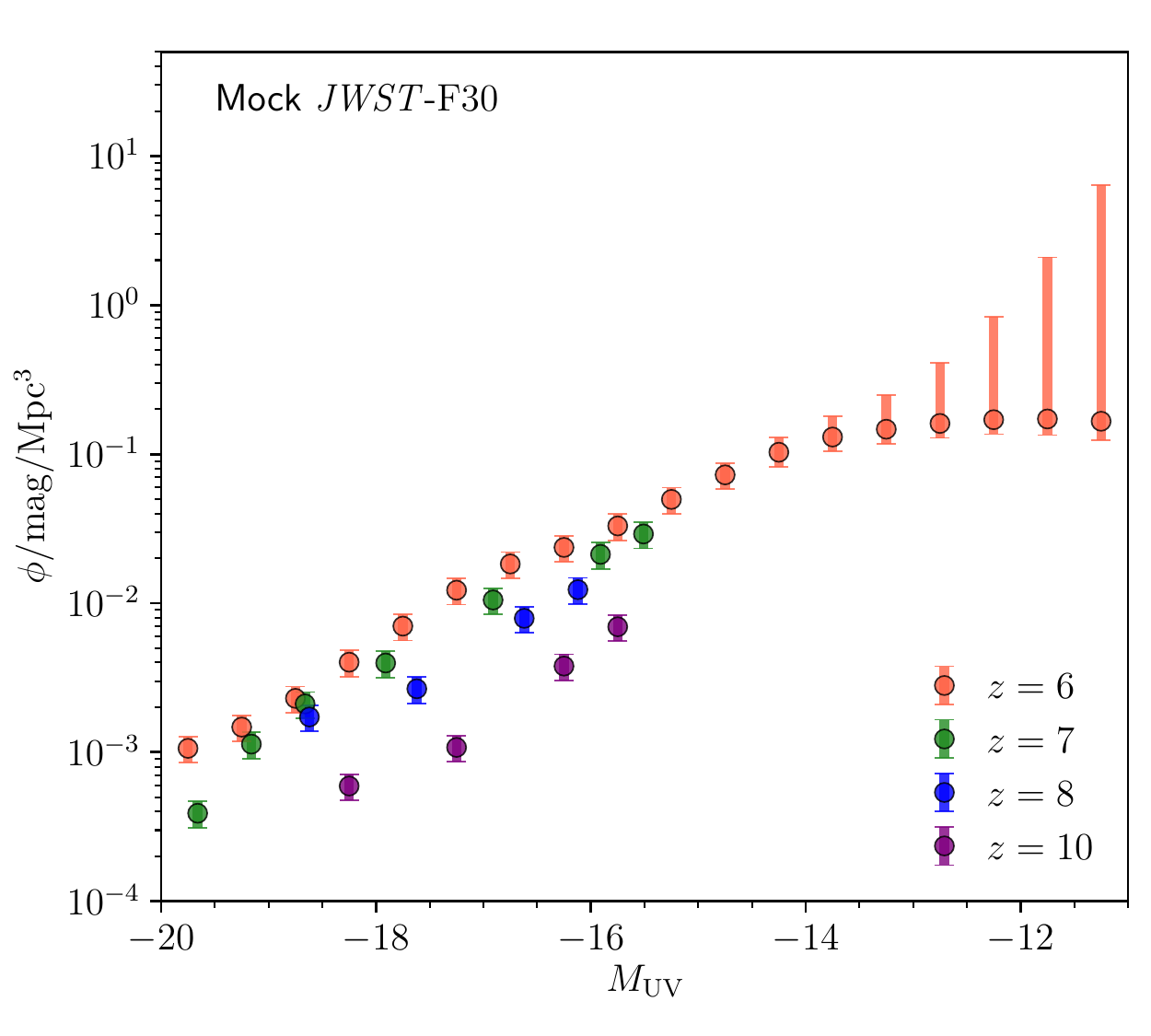}
}
\caption{
  Mock LFs assuming an optimistic error budget, obtained by reducing the fiducial {\it JWST} uncertainties by 30\%.  The intrinsic number densities are taken from the faint turnover model.
  \label{fig:LFs2}
}
\vspace{-0.5\baselineskip}
\end{figure}

Fig. \ref{fig:LFs2} shows the resulting {\it JWST} LFs, and the corresponding parameter constraints are shown in Fig.~\ref{fig:corner_LFs1} with the label {\it JWST}-F30.
The most notable improvement is on ${\rm log}_{10}(M_{\rm turn})$. As evidenced by the 1D PDF, the constraints on the turnover mass are significantly tightened, which is in contrast to the mock {\it JWST}-F LFs which only provide an upper limit. Moreover, the $1\sigma$ fractional uncertainty for $\alpha_{\ast}$ is reduced by $\sim 60$ per cent, compared with the mock {\it JWST}-F LFs.

\subsubsection{Assuming an intrinsic turnover at brighter magnitudes}

\begin{figure*}
\vspace{-1\baselineskip}
{
 \includegraphics[trim={0 5cm 0 0},clip,width=\textwidth]{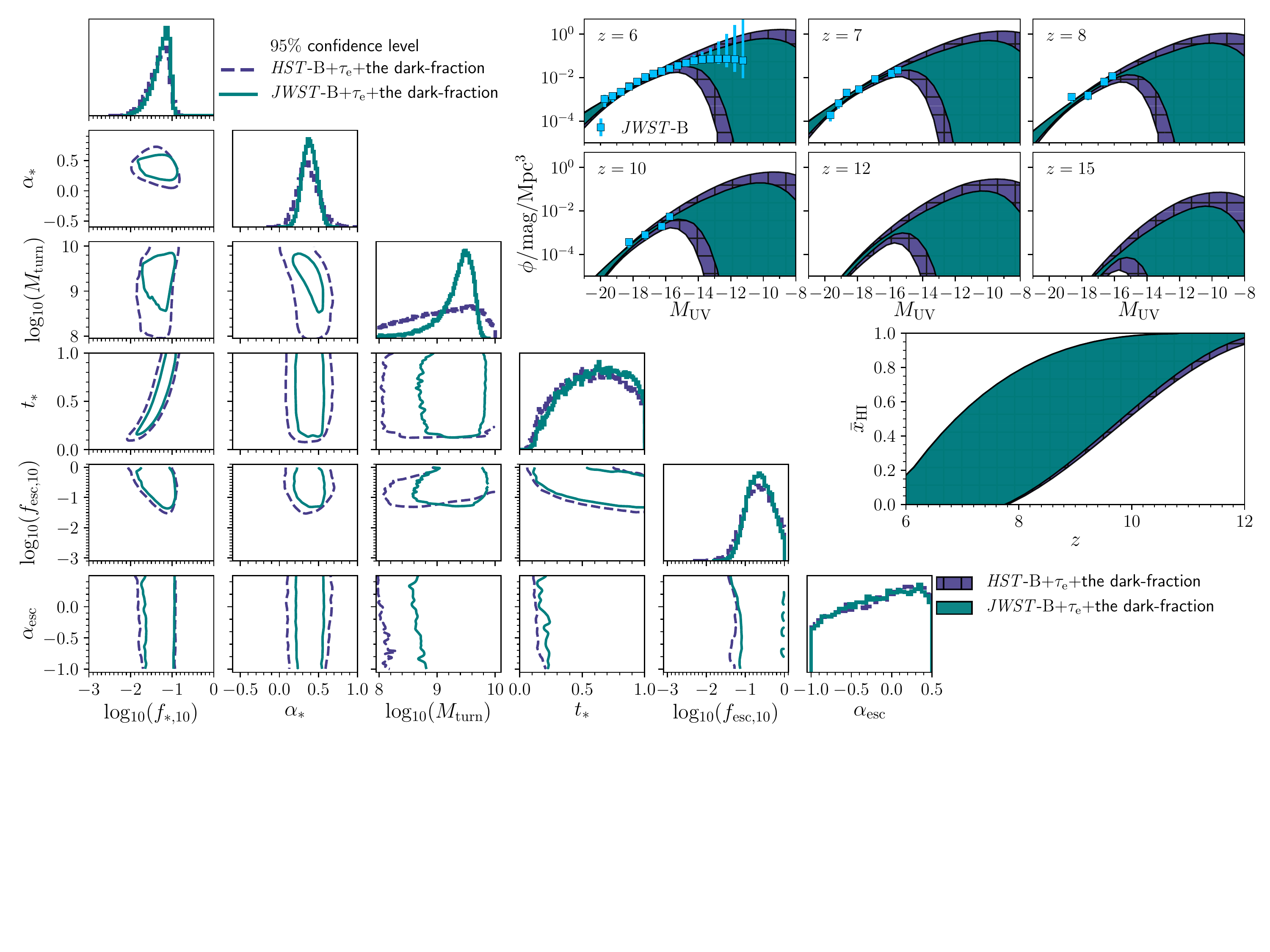}

}
\caption{
 The same as Fig. \ref{fig:corner_LFs1}, but for the mock {\it HST}-B and {\it JWST}-B LFs. Purple dashed lines and Turquoise solid lines represent $95$ per cent confidence levels for constraints using data sets of the mock ${\it HST}$-B and the mock ${\it JWST}$-B, respectively. Shaded regions with the cross hatch (purple, `$+$') and shaded regions (turquoise) represent constraints using the mock ${\it HST}$-B and the mock ${\it JWST}$-F, respectively.
\label{fig:corner_LFs2}
}
\vspace{-0.5\baselineskip}
\end{figure*}

We now show the resulting constraints for the mock LFs with the turnover at brighter magnitudes (i.e., {\it HST}-B and {\it JWST}-B from Fig. 2) in Fig. \ref{fig:corner_LFs2}. Comparing {\it JWST}-B results to those of {\it JWST}-F, we note a large improvement in the inference of the turnover scale.
This is understandable, since the ``-B'' LFs intrinsically turn over at scales which approach the {\it JWST} sensitivity thresholds.
This is reflected also in the recovered luminosity functions (upper right panels in Fig. \ref{fig:corner_LFs2}), which understandably show stronger evidence of a turnover than was the case for  {\it JWST}-F in the previous figure.
Specifically, we recover
${\rm log}_{10}(M_{\rm turn})=9.40^{+0.18}_{-0.36}$ ($1\sigma$).  This fractional uncertainty of $\sim 3$ per cent is comparable to the constrains achievable with the reduced error bar LFs discussed in \S~\ref{sec:result_JWST-F30}.  Therefore, significant improvement in the inference of faint-end galaxy properties is likely with {\it JWST} if {\it either} we are able to better characterize systematic lensing uncertainties than currently possible, {\it or} the intrinsic LFs peak at $M_{\rm UV} \lsim -13$.

\begin{figure*}
\vspace{-1\baselineskip}
{
 \includegraphics[width=\textwidth]{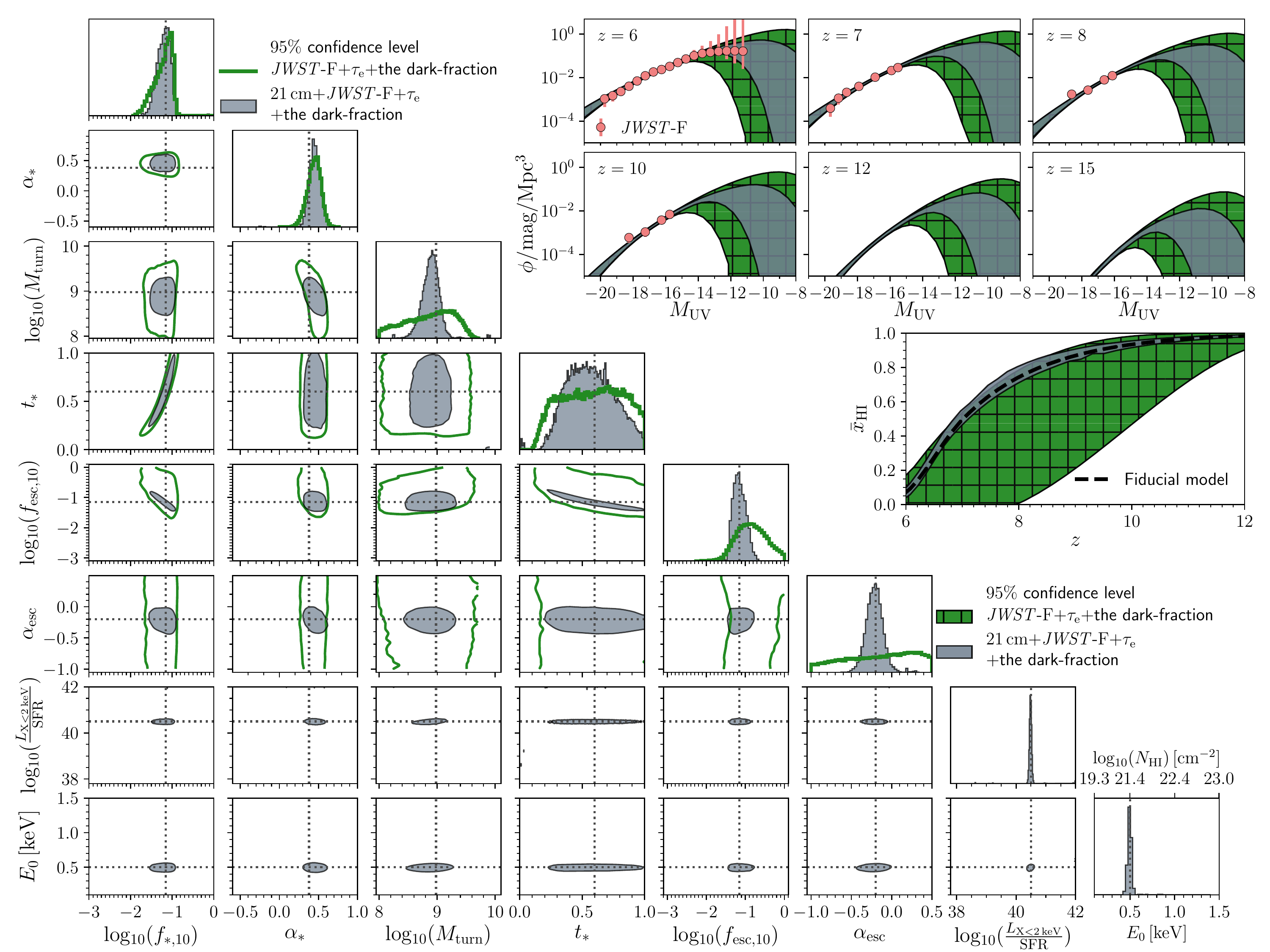}
}
\caption{
  Same as Fig. \ref{fig:corner_LFs1}, but including also constraints available from adding mock 21-cm observations (see legend).
\label{fig:corner}
}
\vspace{-0.5\baselineskip}
\end{figure*}
%
%
\subsection{Combined constraints with LFs and 21-cm signals}\label{sec:result_21cmLFs}

In the previous sections, we saw that if the LFs turn over at $M_{\rm UV} \gsim -12$ (our ``-F'' models), a dramatic improvement in inference using {\it JWST} observations is unlikely, given our fiducial uncertainties.  Here we additionally add mock 21-cm PS observations, to see if parameter inference is improved, for these ``pessimistic'' LFs.  We also extend our parameter space to include the afore-mentioned X-ray parameters, which drive the Epoch of Heating, observable with 21-cm.

The resulting corner plot is shown in Fig. \ref{fig:corner}.  Adding the mock 21-cm observation results in a marked improvement in all parameter constraints, as expected from \cite{Park2019}. We find the 21-cm signal dominates constraints on ${\rm log_{10}}(M_{\rm turn})$, $t_{\ast}$, ${\rm log_{10}}(f_{\rm esc,10})$, $\alpha_{\rm esc}$, ${\rm log_{10}}(L_{\rm X<2\,keV}/{\rm SFR})$ and $E_0$. On the other hand, the LFs dominates constraints on $\alpha_{\ast}$, as evidenced by the almost identical 1D PDFs of $\alpha_{\ast}$ from the mock 21-cm + {\it JWST}-F LFs and from the mock {\it JWST}-F LFs only. Constraints on $f_{\ast, 10}$ are comparably sourced by both observations. In summary, the $1\,\sigma$ fractional uncertainties on our parameters from the combined data sets are [${\rm log_{10}}(f_{\ast,10})$, $\alpha_{\ast}$, ${\rm log_{10}}(f_{\rm esc,10})$, $\alpha_{\rm esc}$, ${\rm log_{10}}(M_{\rm turn})$, $t_{\ast}$, ${\rm log_{10}}(L_{\rm X<2\,keV}/{\rm SFR})$, $E_0$] $=$ ($11$, $15$, $13$, $47$, $1.0$, $29$, $0.1$, $4.6$) per cent.

The most dramatic improvement is seen in the EoR history (middle right panel).  With 21-cm observations, we will know the EoR history to within $\Delta z(\avenf) \lsim 0.1$ $(1 \sigma)$ over most of the EoR.  This is a order of magnitude improvement over our current state of knowledge: $\Delta z(\avenf) \lsim 1$.

%
%
\section{Conclusions}\label{sec:conc}

Next generation observatories will enable us to study the first billion years of our Universe in unprecedented detail.  Foremost among these are 21-cm interferometry with HERA and SKA, and high-$z$ galaxy observations with {\it JWST}.  Here we quantify how observations from these instruments can be used to constrain the astrophysics of high-$z$ galaxies.  For this purpose, we generate mock {\it JWST} LFs, based on two different hydrodynamical cosmological simulations; these have intrinsic LF which turn over at different scales and yet are fully consistent with present-day observations.
Likewise, we generate mock 21-cm power spectra, using the semi-numerical code \cmfast\, combined with a moderate foreground model and 1000h thermal noise with the SKA1-low instrument.  We assume a simple astrophysical model for the high-$z$ galaxy population, in which the star formation rate and ionizing escape fraction are power-law functions of halo mass, and there is an exponential suppression of star forming galaxies below some threshold halo mass.

We find that if the LFs turn over at magnitudes fainter than $M_{\rm UV} \gsim -12$, we must significantly improve on our understanding of systematic lensing uncertainties in order for {\it JWST} LFs to dramatically improve our understanding of the faint galaxies, beyond what we have currently with {\it HST} LFs.  However, if LFs intrinsically turn over at magnitudes brighter than $M_{\rm UV} \lsim -13$, then the turn over scale can be easily recovered to within a few percent, and uncertainties on the star formation rate to halo mass relation can be decreased by $\sim 50\%$.

Additionally including 21-cm observations would improve constraints significantly, even for our most pessimistic {\it JWST} scenario.  The two observations are complementary, with JWST dominating constraints on the star formation rate to halo mass relation, and 21-cm dominating constraints on the ionizing escape fraction, turn over scale, and the EoR history.

\section*{Acknowledgements}
We thank R. Bouwens, S. Finkelstein and P. Oesch for enlightening discussions about approximating {\it JWST}  LF uncertainties.  This project has received funding from the European Research Council (ERC) under the European Union's Horizon 2020 research and innovation program (grant agreement No. 638809 -- AIDA -- PI: Mesinger).  The results presented here reflect the authors' views; the ERC is not responsible for their use. The simulations used to generate the mock LFs were computed as part of the PRACE tier-0 grant GAFFER (project no. 2016163945). We acknowledge PRACE for awarding us access to Curie at GENCI@CEA, France.




\bibliographystyle{mnras}
\bibliography{ref.bib} 







\bsp	
\label{lastpage}
\end{document}